\documentclass[12pt]{article}
\usepackage{epsfig}

\def\beq{\begin{equation}}
\def\eeq#1{\label{#1}\end{equation}}
\def\eeqn{\end{equation}}

\def\beqa{\begin{eqnarray}}
\def\eeqa#1{\label{#1}\end{eqnarray}}
\def\eeqan{\end{eqnarray}}
\def\CR{\nonumber \\ }

\def\leqn#1{(\ref{#1})}

\let\bar=\overbar

\def\Dslash{\not{\hbox{\kern-4pt $D$}}}
\def\dslash{\not{\hbox{\kern-2pt $\del$}}}

\def\BR{\mbox{\rm BR}}

\def\msb{{\bar{\ssstyle M \kern -1pt S}}}

\newcommand{\BABARPubYear}    {02}
\newcommand{\BABARProcNumber} {048}

\input babarsym.tex

\def\pipi     {\ensuremath{\pi\pi}}

\def\hh     {\ensuremath{h^+h^{\prime -}}}

\def\fpm {\ensuremath{f_{\pm}(\deltat)}}
\def\ilam {\ensuremath{{\cal I}m\lambda}}
\def\alam {\ensuremath{\left|\lambda\right|}}

\def\spipi {\ensuremath{S_{\pi\pi}}}
\def\cpipi {\ensuremath{C_{\pi\pi}}}
\def\de {\ensuremath{\Delta E}}

\def\diffD {\ensuremath{\Delta D}}

\def\Btag {\ensuremath{B_{\rm tag}}}

\def\Bflav {\ensuremath{B_{\rm flav}}}

\def\ttag {\ensuremath{t_{\rm tag}}}

\def\Title#1{\begin{center} {\Large {\bf #1} } \end{center}}

\begin{document}
\begin{flushright}
\babar-PROC-\BABARPubYear/\BABARProcNumber \\
\end{flushright}

\Title{\boldmath Measurements of branching fractions and \CP-violating
asymmetries in $\Bz\to\pip\pim,\, \Kp\pim,\, \Kp\Km$ decays}

\bigskip\bigskip


\begin{raggedright}  
{\it Paul D. Dauncey \index{Dauncey, P.D.} (representing the \babar\ Collaboration)\\
Blackett Laboratory\\
Imperial College\\
Prince Consort Road\\
London SW7 2BW, UK}
\end{raggedright}
\bigskip

\begin{center}
{Presented at ``Flavor Physics and \CP Violation'' (FPCP),\\
16-18 May 2002, University of Pennsylvania, Philadelphia, USA.}
\end{center}

\section{Introduction}
Recent measurements of the $\CP$-violating asymmetry parameter $\stwob$ by the
\babar~\cite{BaBarSin2betaObs} \index{BaBar} and 
BELLE~\cite{BelleSin2betaObs} \index{Belle} collaborations
established \CP violation in the $\Bz$ system.  These measurements, as well as 
updated preliminary results~\cite{BaBarSin2betaM02, BelleSin2betaM02}, are 
consistent with the Standard Model expectation based on measurements and 
theoretical estimates of the elements of the 
Cabibbo-Kobayashi-Maskawa~\cite{CKM} (CKM) quark-mixing matrix.

The study of $B$ decays to charmless hadronic two-body final states will
yield important information about the remaining angles ($\alpha$ and $\gamma$) 
of the Unitarity Triangle.  In the Standard Model, the time-dependent 
\CP-violating asymmetry in the decay $\Bz\to\pip\pim$ is related to the 
angle $\alpha$, \index{alpha}
and ratios of branching fractions for various $\pi\pi$ 
and $K\pi$ decay modes are sensitive to the angle $\gamma$.
We previously reported measurements of branching fractions~\cite{twobodyPRL}
and \CP-violating asymmetries~\cite{BaBarSin2alpha} \index{CP violation}
in $\Bz\to\pip\pim$, $\Kp\pim$, and $\Kp\Km$ 
decays. (Unless explicitly stated, charge conjugate
decay modes are assumed throughout this paper.)
In this paper, we present a preliminary update of these results
using a sample of $60$ million $\BB$ pairs.

We reconstruct a sample of $B$ mesons ($B_{\rm rec}$) decaying to the $\hh$ 
final state, where $h$ and $h^{\prime}$ refer to $\pi$ or $K$, and examine the 
remaining charged particles in each event to ``tag'' the flavor of the other 
$B$ meson (\Btag).  The decay rate distribution $f_+\,(f_-)$ when 
$\hh = \pip\pim$
and $\Btag = \Bz\,(\Bzb)$ is given by
\beq
\fpm = \frac{e^{-\left|\deltat\right|/\tau}}{4\tau} 
\left[1\pm \spipi\sin(\deltamd\deltat) \mp \cpipi\cos(\deltamd\deltat)\right],
\eeq{fplusminus}
where $\tau$ is the mean $\Bz$ lifetime, $\deltamd$ is the 
eigenstate mass difference, 
and $\deltat = t_{\rm rec} - \ttag$ is the time between the $B_{\rm rec}$ and 
\Btag\ decays.  The \CP-violating parameters $\spipi$ and $\cpipi$ 
are defined in terms of a complex parameter $\lambda$ as
\beq
\spipi = \frac{2\,\ilam}{1+\alam^2} \qquad {\rm and} \qquad 
\cpipi = \frac{1-\alam^2}{1+\alam^2}.
\eeq{SandCdef}
If the decay proceeds purely through the $b\to uW^-$ tree process,
then $\lambda$ is given in terms of the CKM elements $V_{\rm ij}$ by
\beq
\lambda(B\to\pip\pim) 
= \left(\frac{V_{\rm tb}^*V_{\rm td}}{V_{\rm tb}V_{\rm td}^*}\right)
\left(\frac{V_{\rm ud}^*V_{\rm ub}}{V_{\rm ud}V_{\rm ub}^*}\right).
\eeqn
In this case, $\cpipi = 0$ and $\spipi = \stwoa$, where 
$\alpha \equiv 
\arg\left[-V_{\rm td}V_{\rm tb}^*/V_{\rm ud}V_{\rm ub}^*\right]$.  
In general, the $b\to dg$ penguin amplitude modifies 
both the magnitude and phase
of $\lambda$, so that $\cpipi \ne 0$ and 
$\spipi = \sqrt{1 - \cpipi^2}\sin{2\alpha_{\rm eff}}$, 
where $\alpha_{\rm eff}$ depends on the magnitudes and relative strong and weak
phases of the tree and penguin amplitudes.
Several approaches have been proposed 
to obtain information on $\alpha$ in the presence of 
penguins~\cite{alphafrompenguins}.

\section{The \babar\ detector and dataset}
The data sample used in this analysis consists of $55.6\invfb$, corresponding
to $60.2\pm 0.7$ million $\BB$ pairs, collected on the $\Y4S$ resonance
with the \babar\ detector at the SLAC PEP-II storage ring between 
October 1999 and December 2001.  
Equal branching fractions for \upsbzbz\ and $\Bu\Bub$ are assumed.  

A detailed description of the \babar\ detector 
is presented in~\cite{ref:babar}.  Charged particle (track) momenta are 
measured in a tracking system consisting of a 5-layer double-sided silicon 
vertex tracker (SVT) and a 40-layer drift chamber (DCH) filled with a gas 
mixture of helium and isobutane.  The SVT and DCH operate within a 
$1.5\,{\rm T}$ superconducting solenoidal magnet.  
Photons are detected in an electromagnetic calorimeter (EMC) consisting of 
6580 CsI(Tl) crystals arranged in barrel and forward endcap subdetectors.  
The flux return for the solenoid is composed of multiple layers of iron and 
resistive plate chambers for the identification of muons and long-lived neutral 
hadrons.
Tracks from the $B_{\rm rec}$ decay are identified as pions or kaons by the 
Cherenkov angle $\theta_c$ measured with a detector of internally reflected 
Cherenkov light (DIRC).

\section{Analysis method}
Event selection is identical to that described in~\cite{BaBarSin2alpha}.
Candidate $\B_{\rm rec}$ decays are reconstructed from pairs of 
oppositely-charged 
tracks forming a good quality vertex, where the $B_{\rm rec}$ 
four-vector is calculated
assuming the pion mass for both tracks.  We require each track 
to have an associated $\theta_c$ measurement with a minimum of six Cherenkov 
photons above background, where the average is approximately 30 
for both pions and 
kaons.  Protons are rejected based on $\theta_c$ and electrons 
are rejected based on $dE/dx$ measurements in the tracking system,
shower shape 
in the EMC, and the ratio of shower energy and track momentum.
Background from the reaction 
$\epem\to q\bar{q}\; (q=u,d,s,c)$ is suppressed by removing jet-like events 
from the sample: we define the center-of-mass (CM) angle 
$\theta_S$ between the sphericity 
axes of the $B$ candidate and the remaining tracks and photons in the event, 
and require 
$\left|\cos{\theta_S}\right|<0.8$, which removes $83\%$ of the background.  
The total efficiency for signal events of the above selection is 
approximately $38\%$.

Signal decays are identified kinematically using two variables.
We define a beam-energy substituted mass 
$\mes = \sqrt{E^2_{\rm b}- {\mathbf {p}}_B^2}$, where the $B$ candidate energy 
is defined as $E_{\rm b} =(s/2 + {\mathbf {p}}_i\cdot {\mathbf {p}}_B)/E_i$, 
$\sqrt{s}$ and $E_i$ are the total energies of the \epem\ system in the
CM and laboratory frames, respectively, and ${\mathbf {p}}_i$ and 
${\mathbf {p}}_B$ are the momentum vectors in the laboratory frame of 
the \epem\ system and the $B_{\rm rec}$ candidate, respectively.  
Signal events are Gaussian distributed in $\mes$ with a mean near the $B$ mass 
and a resolution of $2.6\mevcc$, dominated by the beam energy spread.  
The background shape is parameterized by a threshold function~\cite{ARGUS}
with a fixed endpoint given by the average beam energy.

We define a second kinematic variable $\de$ as the difference between the 
energy of the $B_{\rm rec}$ candidate in the CM frame and $\sqrt{s}/2$.  
Signal $\pi\pi$ decays are Gaussian distributed with a mean value near zero.  
For decays with one\,(two) kaons, the distribution is shifted relative to 
$\pi\pi$ on average by $-45\mev$ ($-91\mev$), respectively, where the exact 
separation depends on the laboratory momentum of the kaon(s).  The resolution 
on $\de$ is approximately $26\mev$ and is validated in large samples of fully 
reconstructed $B$ decays.  The background is parameterized by a quadratic 
function.  

Candidate $\hh$ pairs selected in the region $5.2 < \mes < 5.3\gevcc$ 
and $\left|\de\right|<0.15\gev$ are used to extract yields and \CP-violating 
asymmetries with an unbinned maximum likelihood fit.
The total number of events in 
the fit region satisfying all of the above criteria is $17585$.

To determine the flavor of the \Btag\ meson we use the same $B$-tagging 
algorithm used in the \babar\ $\stwob$ analysis~\cite{ref:sin2betaPRD}.
The algorithm relies on the correlation between the flavor of the $b$ quark
and the charge of the remaining tracks in the event after removal of the
$B_{\rm rec}$ candidate.
We define five mutually exclusive tagging 
categories: {\tt Lepton}, {\tt Kaon}, {\tt NT1}, {\tt NT2}, and 
{\tt Untagged}.  {\tt Lepton} tags rely on primary electrons and muons
from semileptonic 
$B$ decays, while {\tt Kaon} tags exploit the correlation in the process
$b\to c\to s$ 
between the net kaon charge and the charge of the $b$ quark.  
The {\tt NT1}\,(more certain tags) and {\tt NT2}\,(less certain tags)
categories are derived from a neural network that is sensitive 
to charge correlations between the parent \B\ and unidentified leptons
and kaons, soft pions, or the charge and momentum of the track with
the highest CM momentum.
The addition of {\tt Untagged} events improves the signal yield estimates
and provides a larger sample for determining background shape parameters
directly in the maximum likelihood fit.

The quality of tagging is expressed in terms of the effective efficiency 
$Q = \sum_c \epsilon_c D_c^2$, where $\epsilon_c$ is the fraction of events
tagged in 
category $c$ and the dilution $D_c = 1-2w_c$ is related to the mistag
fraction $w_c$.  
Table~\ref{tab:tagging} summarizes the tagging performance in a data
sample \Bflav\ of fully reconstructed neutral $B$ decays into
$D^{(*)-}h^+\,(h^+ = \pip, \rho^+, a_1^+)$ 
and $\jpsi K^{*0}\,(K^{*0}\to\Kp\pim)$ flavor eigenstates.
We use the same tagging efficiencies and dilutions for signal 
$\pi\pi$, $K\pi$, and $KK$ decays.
Separate background efficiencies for each species are determined
simultaneously with $\spipi$ and $\cpipi$ in the maximum likelihood fit.
\begin{table}[t]
\begin{center}
\begin{tabular}{l|cccc}
Category & $\epsilon\,(\%)$ & $D\,(\%)$ & $\diffD\,(\%)$ & $Q\,(\%)$ \rule[-2mm]{0mm}{6mm} \\\hline
{\tt Lepton}   & $11.1\pm 0.2$ & $82.8 \pm 1.8$ & $-1.2  \pm 3.0$ & $\phantom{1}7.6\pm  0.4$ \rule[-1.5mm]{0mm}{5mm}\\
{\tt Kaon}     & $34.7\pm 0.4$ & $63.8 \pm 1.4$ & $\phantom{-}1.8  \pm 2.1$ & $14.1\pm 0.6$ \rule[-1.5mm]{0mm}{4mm}\\
{\tt NT1}      & $\phantom{1}7.6 \pm 0.2$ & $56.0 \pm 3.0$ & $-2.7  \pm 4.7$ & $\phantom{1}2.4\pm  0.3$ \rule[-1.5mm]{0mm}{1.5mm}\\
{\tt NT2}      & $14.0\pm 0.3$ & $25.4 \pm 2.6$ & $\phantom{-}9.4  \pm 3.8$ & $\phantom{1}0.9\pm  0.2$ \rule[-1.5mm]{0mm}{1.5mm}\\
{\tt Untagged} & $32.6\pm 0.5$ & -- 	  & --            & --          \rule[-1.5mm]{0mm}{1.5mm}\\ \hline
Total $Q$ & & & & $25.0\pm 0.8$ \rule[-2mm]{0mm}{6mm} \\\hline
\end{tabular}
\caption{Tagging efficiency $\epsilon$, average dilution 
$D = 1/2\left(D_{\Bz} + D_{\Bzb}\right)$, dilution difference
$\diffD = D_{\Bz} - D_{\Bzb}$, and effective tagging efficiency
$Q$ for signal events in each tagging category.
The values are measured in the \Bflav\ sample.}
\label{tab:tagging}
\end{center}
\end{table}

The time difference $\deltat$ is obtained from the measured distance between 
the $z$ positions of the $B_{\rm rec}$ and $\Btag$ decay vertices and the 
known boost 
of the $\epem$ system.  The $z$ position of the \Btag\ vertex is determined 
with an iterative procedure that removes tracks with a large contribution to 
the total $\chi^2$.  An additional 
constraint is constructed from the three-momentum and vertex position of the 
$B_{\rm rec}$ candidate, and the average $\epem$ interaction point and boost.
For $99.5\%$
of candidates with a reconstructed vertex, the rms $\deltaz$ resolution is 
$180\mum\,(1.1\ps)$.  We require $\left|\deltat\right|<20\ps$ and 
$\sigma_{\deltat} < 2.5\ps$, where $\sigma_{\deltat}$ is the error on 
$\deltat$.  
The resolution function for signal candidates is a sum of three Gaussians,
identical 
to the one described in~\cite{BaBarSin2betaM02}, with parameters determined
from a 
fit to the \Bflav\ sample (including events in all five tagging categories).  
The background $\deltat$ distribution is parameterized as the sum of an 
exponential
convolved with a Gaussian, and two additional Gaussians to account for tails.  
A common parameterization is used for all tagging categories, and 
the parameters are determined simultaneously with the \CP parameters in the 
maximum 
likelihood fit.  We find that $86\%$ of background events are described by 
an effective 
lifetime of about $0.6\ps$, while tails are described by $12\, (2)\%$ of 
events with
a resolution of approximately $2\,(8)\ps$.

Discrimination of signal from light-quark background is enhanced by the use 
of a Fisher discriminant ${\cal F}$~\cite{twobodyPRL}.  The discriminating
variables are constructed from the scalar sum of the CM
momenta of all tracks and photons 
(excluding tracks from the $B_{\rm rec}$ candidate) entering nine two-sided 
$10$-degree concentric 
cones centered on the thrust axis of the $B_{\rm rec}$ candidate.  
The distribution of ${\cal F}$ for signal events is parameterized as a single 
Gaussian, 
with parameters determined from Monte Carlo simulated decays and validated 
with $\Bub\to\Dz\pim$ decays reconstructed in data.
The background shape is parameterized as 
the sum of two Gaussians, with parameters determined directly in the 
maximum likelihood fit.

\begin{figure}[htb]
\begin{center}
\epsfig{file=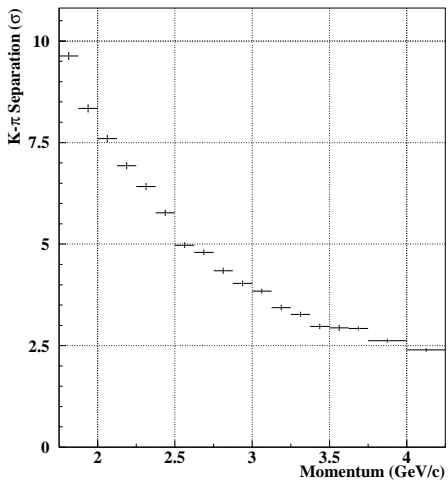,height=2.5in}
\caption{Variation of the separation between the kaon and pion Cherenkov 
angles with momentum, as obtained from a control sample of
$D^{*+}\to \Dz\pip$, $\Dz\to\Km\pip$ decays reconstructed in data.}
\label{fig:dirc}
\end{center}
\end{figure}

Identification of $\hh$ tracks as pions or kaons is accomplished with the 
Cherenkov angle measurement from the DIRC.  We construct Gaussian probability
density functions 
(PDFs) from the difference between measured and expected values of 
$\theta_c$ for the pion or 
kaon hypothesis, normalized by the resolution.  The DIRC performance 
is parameterized using a sample of $D^{*+}\to\Dz\pip$, $\Dz\to \Km\pip$ 
decays, reconstructed in data.
Figure~\ref{fig:dirc} shows the typical separation between pions and kaons,
which varies from $8\sigma$ at momenta of
$2\gevc$ to $2.5\sigma$ at $4\gevc$, where $\sigma$ 
is the average resolution of $\theta_c$.

\section{Results}
We use unbinned extended maximum likelihood fits to extract yields and 
$\CP$ parameters
from the $B_{\rm rec}$ sample.  The likelihood for candidate $j$ tagged in 
category $c$ is obtained by summing the product of event yield $n_{i}$, 
tagging efficiency $\epsilon_{i,c}$,
and probability ${\cal P}_{i,c}$ over the eight possible signal and 
background hypotheses $i$
(referring to $\pi\pi$, $\Kp\pim$, $\Km\pip$, and $KK$ decays),
\beq
{\cal L}_c = \exp{\left(-\sum_{i}n_i\epsilon_{i,c}\right)}
\prod_{j}\left[\sum_{i}n_i\epsilon_{i,c}{\cal P}_{i,c}(\vec{x}_j;
\vec{\alpha}_i)\right].
\eeqn
For the $K^{\mp}\pi^{\pm}$ components, the yield is parameterized as 
$n_i = N_{K\pi}\left(1 \pm {\cal A}_{K\pi}\right)/2$, where 
$N_{K\pi} = N_{\Km\pip} + N_{\Kp\pim}$ and 
${\cal A}_{K\pi}\equiv (N_{\Km\pip} - 
N_{\Kp\pim})/(N_{\Km\pip} + N_{\Kp\pim})$.
The probabilities ${\cal P}_{i,c}$ are evaluated as the product of PDFs 
for each of the independent variables 
$\vec{x}_j = \left\{\mes, \de, {\cal F}, \theta_c^+, \theta_c^-, 
\deltat\right\}$, 
where $\theta_c^+$ and $\theta_c^-$ are the Cherenkov angles 
for the positively and 
negatively charged tracks.  We use identical PDFs for $\theta_c^+$ and
$\theta_c^-$.  The total likelihood ${\cal L}$ is the product of likelihoods 
for each tagging category and the free parameters are determined by 
minimizing the quantity $-\ln{\cal L}$.

\subsection{Time-independent fit}
In order to minimize the
systematic error on the branching fraction measurements, we perform an
initial fit without tagging or $\deltat$ information.
A total of $16$ parameters are
varied in the fit, including signal and background yields (6 parameters)
and asymmetries (2), 
and parameters for the background shapes in $\mes$ (1), $\de$ (2),
and ${\cal F}$ (5).  

Table~\ref{tab:BR} summarizes results for total efficiencies, signal yields
and branching fractions.
The upper limit on the signal yield for \
$\Bz\to\Kp\Km$ is given by the value of $n^0$ for which 
$\int_0^{n^0} {\cal L}_{\rm max}\,dn/\int_0^\infty 
{\cal L}_{\rm max}\,dn = 0.90$, 
where ${\cal L}_{\rm max}$ is the likelihood as a function of $n$, 
maximized with respect to the remaining fit parameters.
The branching fraction upper limit is 
calculated by increasing the signal yield upper limit and 
reducing the efficiency by their 
respective systematic errors.
\begin{table}[t]
\begin{center}
\begin{tabular}{l|ccc} 
Mode  &  $\epsilon_T$ (\%) & $N_S$ (Events) & \BR\ ($10^{-6}$) \\ \hline
$\pip\pim$ & $38.5\pm 0.7$  & $124^{+16\, +7}_{-15\, -9}$ & $5.4\pm 0.7\pm 0.4$ \\
$\Kp\pim$ & $37.6\pm 0.7$ & $403\pm 24\pm 15$ & $17.8\pm 1.1\pm 0.8$ \\
$\Kp \Km$  & $36.7\pm 0.7$ & $<16$ ($90\%$ C.L.) & $<1.1$ ($90\%$ C.L.) \\ \hline
\end{tabular}
\caption{Summary of results for total detection efficiencies $\epsilon_T$,
fitted signal yields $N_S$ and measured branching fractions \BR.}
\label{tab:BR}
\end{center}
\end{table}
The fit result for the $K\pi$ charge asymmetry ${\cal A}_{K\pi}$ is
\beq
{\cal A}_{K\pi} = -0.05 \pm 0.06 \pm 0.01, \qquad
90\%\ {\rm C.L.}\ -0.14 < {\cal A}_{K\pi} < 0.05.
\eeqn
The statistical and systematic errors on ${\cal A}_{K\pi}$ are 
added in quadrature when calculating the $90\%$ confidence level (C.L.).

The dominant systematic error on the branching fraction measurements 
is due to uncertainty in the shape of the $\theta_c$ PDF,
while the dominant error on 
${\cal A}_{K\pi}$ is due to possible charge bias in track and $\theta_c$ 
reconstruction.
All measurements are consistent with our previous results 
reported in~\cite{twobodyPRL}.

\begin{figure}[htb]
\begin{center}
\epsfig{file=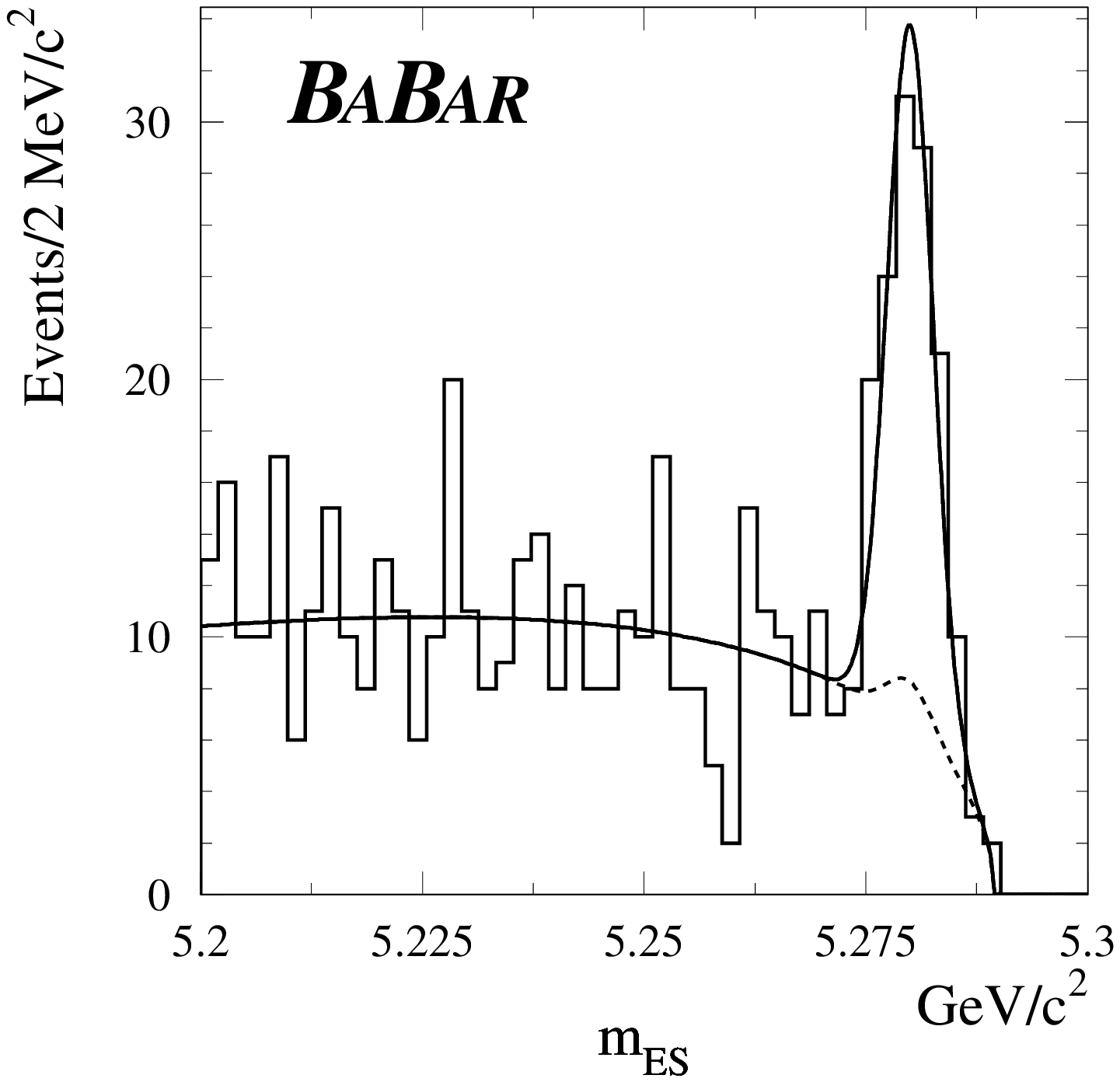,height=2.0in}
\epsfig{file=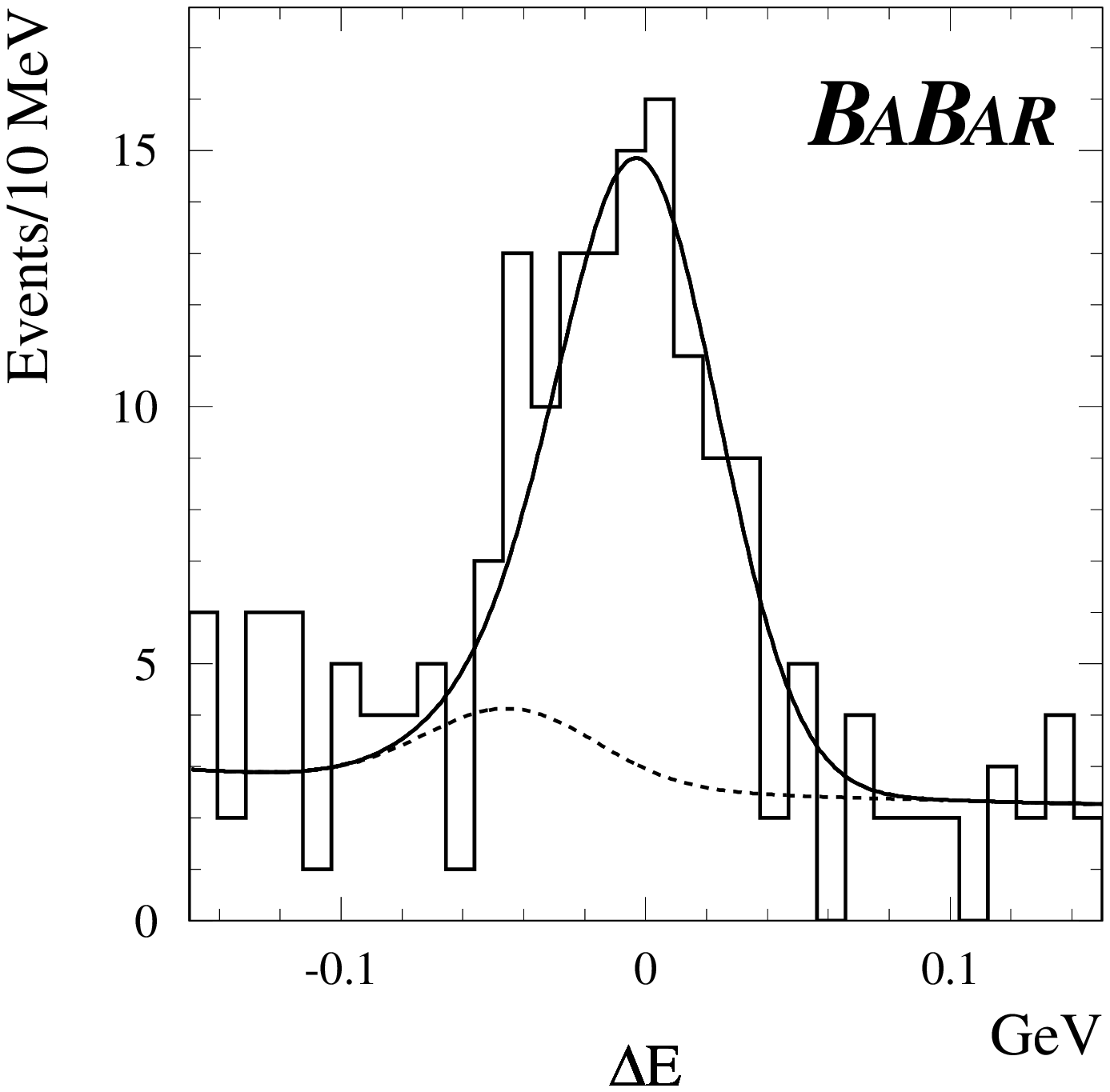,height=2.0in}
\epsfig{file=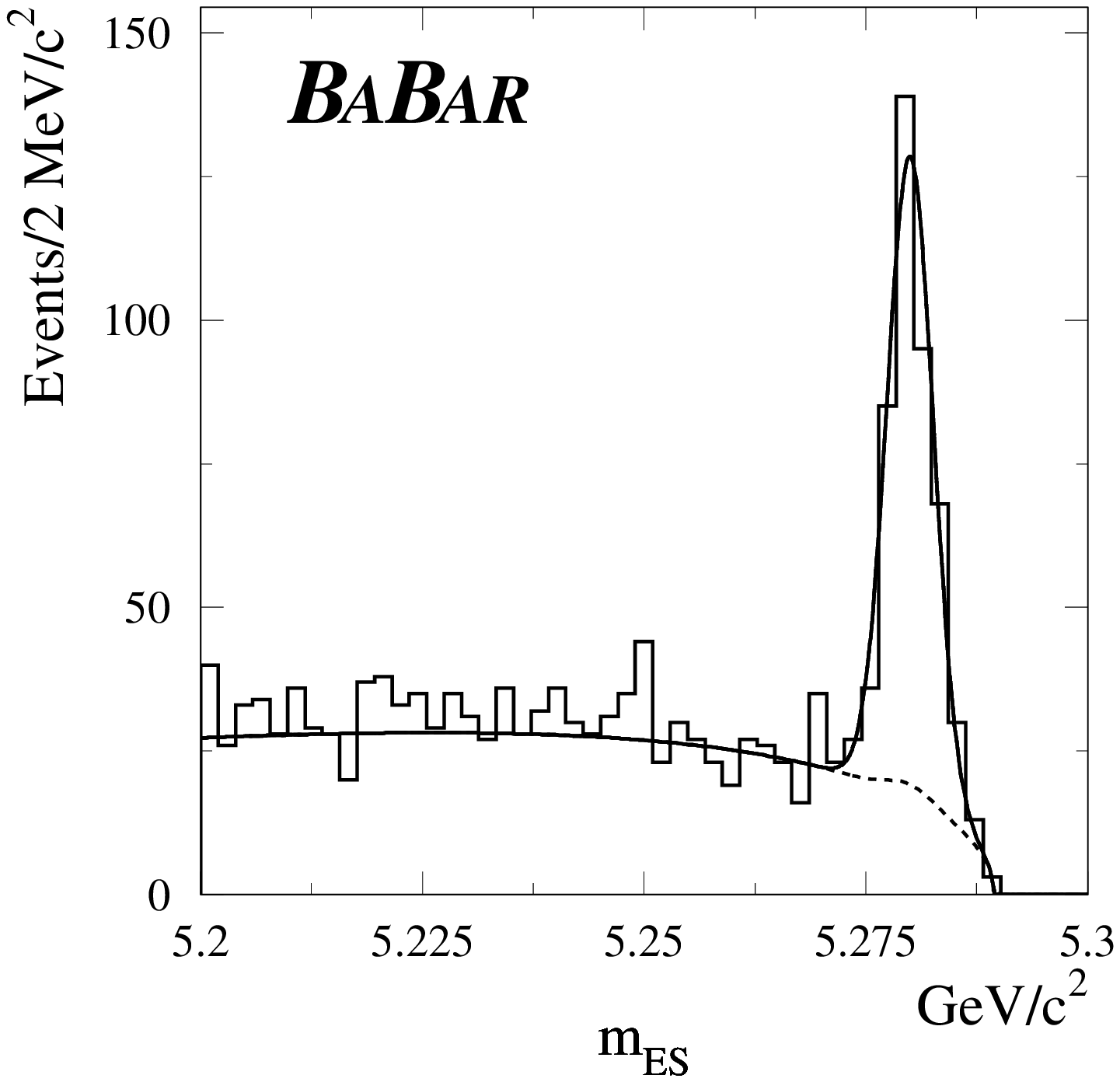,height=2.0in}
\epsfig{file=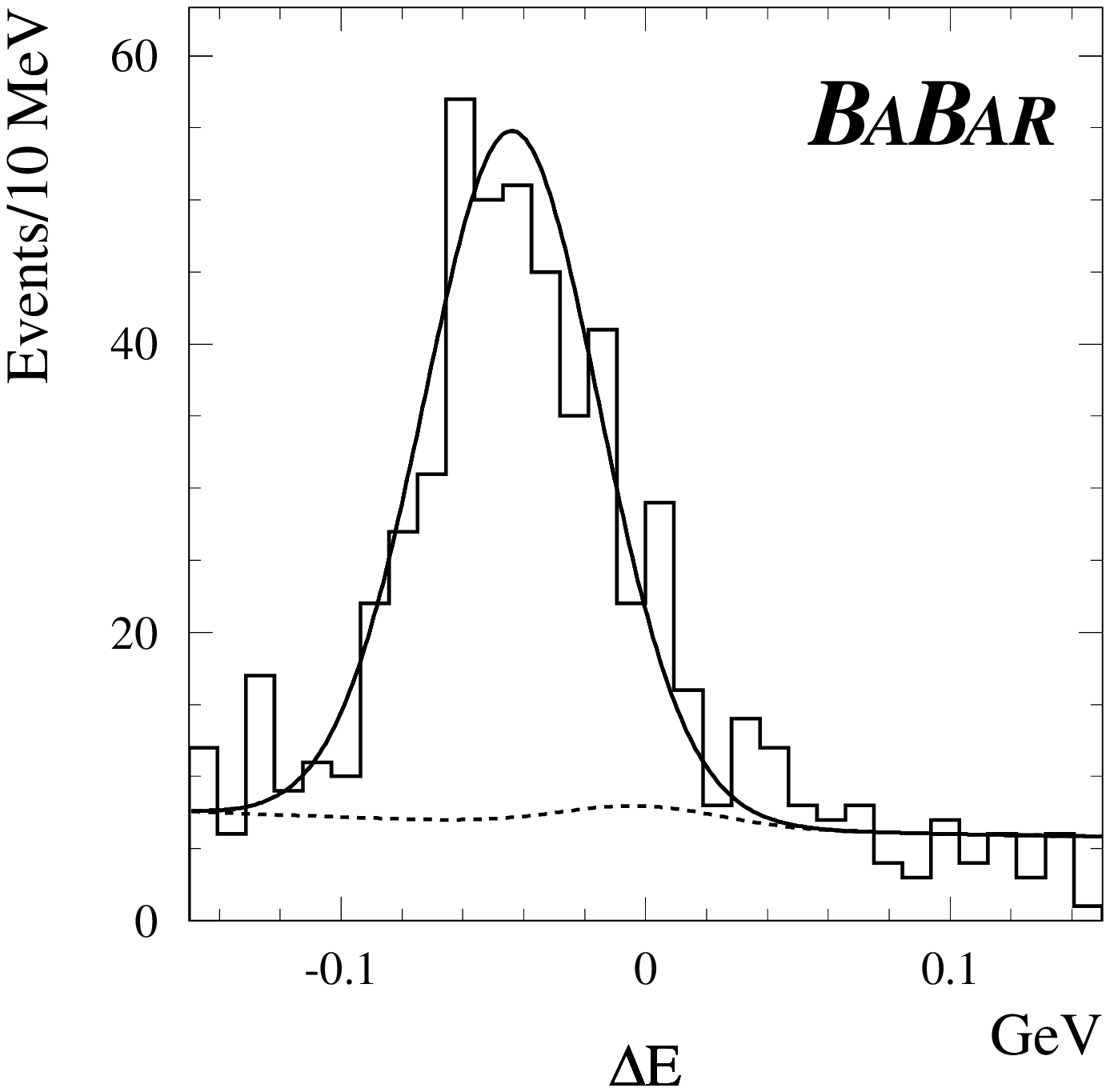,height=2.0in}
\caption{Distributions of $\mes$ (left) and $\de$ (right)
for events enhanced in signal $\pi\pi$ (top) and $K\pi$ (bottom)
decays based on the likelihood ratio selection described in the text.  
Solid curves represent projections of the maximum likelihood fit result 
after accounting for the efficiency of the additional selection, 
while dashed curves represent $q\bar{q}$ and $\pi\pi\leftrightarrow K\pi$ 
cross-feed background.}
\label{fig:prplots}
\end{center}
\end{figure}

Figure~\ref{fig:prplots} shows distributions of $\mes$ and $\de$ after a cut
on likelihood ratios.
We define ${\cal R}_{\rm sig} = \sum_s{n_s{\cal P}_s}/\sum_i{n_i{\cal P}_i}$ 
and 
${\cal R}_k = n_k{\cal P}_k/\sum_s{n_s{\cal P}_s}$, where $\sum_s\, (\sum_i)$
indicates a sum 
over signal\,(all) hypotheses, and ${\cal P}_k$ indicates the probability
for signal
hypothesis $k$.  The probabilities include the PDFs for $\theta_c$,
${\cal F}$, and 
$\mes\,(\Delta E)$ when plotting $\Delta E\,(\mes)$.  The selection is defined
by optimizing the signal significance with respect to ${\cal R}_{\rm sig}$ 
and ${\cal R}_k$.
The solid curve in each plot represents the fit projection after correcting
for the
efficiency of the additional selection (approximately $67\%$ for $\pi\pi$ and 
$88\%$ for $K\pi$).  

\subsection{Time-dependent fit}
The time-dependent \CP asymmetries $\spipi$ and $\cpipi$ are determined
from a second fit 
including tagging and $\deltat$ information, with the yields and
${\cal A}_{K\pi}$ fixed to 
the results of the first fit.  
The $\deltat$ PDF for signal $\pip\pim$ decays is given by 
Eq.~\leqn{fplusminus}, 
modified to
include the dilution and dilution difference for each tagging category,
and convolved with 
the signal resolution function.
The $\deltat$ PDF for signal $K\pi$ events takes into 
account $\Bz$--$\Bzb$ mixing, depending on the charge of the kaon 
and the flavor of $\Btag$.  
We parameterize the $\deltat$ distribution in $\Bz\to\Kp\Km$ decays 
as an exponential 
convolved with the resolution function.

A total of $34$ parameters are varied in the fit,
including the values of $\spipi$ and
$\cpipi$, separate background tagging efficiencies for 
$\pi\pi$, $K\pi$, and $KK$ events $(12)$,
parameters for the background $\deltat$ resolution function $(8)$,
and parameters for the background
shapes in $\mes$ $(5)$,  $\de$ $(2)$, and ${\cal F}$ $(5)$.
The signal tagging efficiencies and dilutions are fixed to the values 
in Table~\ref{tab:tagging},
while $\tau$ and $\deltamd$ are fixed to their PDG values~\cite{PDG2000}.
For each parameter, we also calculate the $90\%$ C.L. interval
taking into account the systematic error. The fit yields
\beqa
\spipi = -0.01\pm 0.37 \pm 0.07,\qquad 
90\%\ {\rm C.L.}\ -0.66 < \spipi < 0.62,\CR
\cpipi = -0.02\pm 0.29 \pm 0.07,\qquad 
90\%\ {\rm C.L.}\ -0.54 < \cpipi < 0.48,
\eeqan
and the correlation between $\spipi$ and $\cpipi$ is $-13\%$.  

Systematic uncertainties on $\spipi$ and $\cpipi$ are dominated by 
the uncertainty on the shape of the $\theta_c$ PDF.
Since we measure asymmetries near zero, multiplicative systematic errors
have also been evaluated $(0.05)$.
We sum in quadrature multiplicative errors, evaluated at
one standard deviation, with the additive systematic uncertainties.

\begin{figure}[htb]
\begin{center}
\epsfig{file=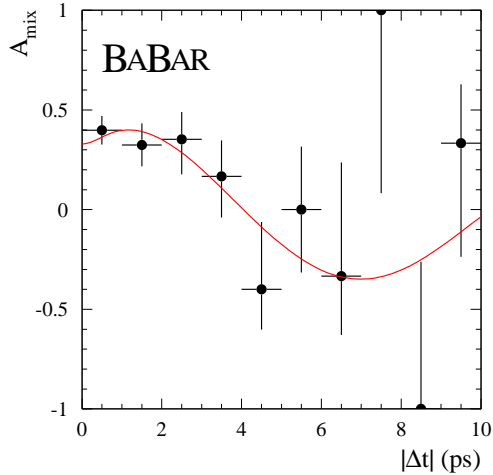,height=2.5in}
\caption{The asymmetry ${\cal A}_{\rm mix}$ between unmixed and mixed events
in a sample enhanced in $K\pi$ decays.  The curve indicates the 
expected oscillation corresponding to $\deltamd = 0.517\,\hbar \ps^{-1}$.
The dilution from $q\bar{q}$ events is evident in the reduced amplitude near
$\left|\deltat\right| = 0$.}
\label{fig:mixing}
\end{center}
\end{figure}

To validate the analysis technique, we measure $\tau$ and $\deltamd$ 
in the $B_{\rm rec}$ sample
and find $\tau = (1.66\pm 0.09)\ps$ and 
$\deltamd = (0.517\pm 0.062)\hbar \ps^{-1}$.  
Figure~\ref{fig:mixing} shows the asymmetry 
${\cal A}_{\rm mix} = (N_{\rm unmixed} - 
N_{\rm mixed})/(N_{\rm unmixed}+N_{\rm mixed})$
in a sample of events enhanced in $B\to K\pi$ decays.
The curve shows the expected 
oscillation given the value of $\deltamd$ measured in the full sample.

\begin{figure}[htb]
\begin{center}
\epsfig{file=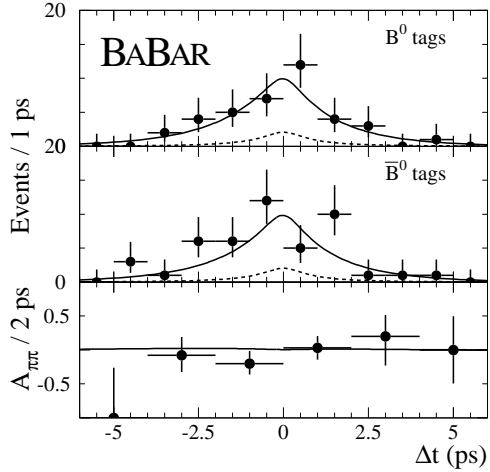,height=2.5in}
\caption{Distributions of $\deltat$ for events enhanced in 
signal $\pi\pi$ decays based on the likelihood ratio selection 
described in the 
text.  The top two plots show events (points with errors) with $\Btag=\Bz$ or 
$\Bzb$.  Solid curves represent projections of the maximum 
likelihood fit, dashed curves represent the sum of $q\bar{q}$ and $K\pi$ 
background events. The bottom plot shows ${\cal A}_{\pi\pi}(\deltat)$ for data 
(points with errors) and the fit projection.}
\label{fig:dtplot}
\end{center}
\end{figure}

For tagged events enhanced in signal $\pi\pi$ decays,
Figure~\ref{fig:dtplot} shows the $\deltat$ distributions and the 
asymmetry ${\cal A}_{\pi\pi}(\deltat) = 
[N_{\Bz}(\deltat) - N_{\Bzb}(\deltat)]/[N_{\Bz}(\deltat) + N_{\Bzb}(\deltat)]$.
The selection procedure is the
same as for Figure~\ref{fig:prplots},
with the likelihoods defined including the PDFs for 
$\theta_c$, ${\cal F}$, $\mes$, and $\de$.

\section{Summary}
In summary, we have presented updated preliminary
measurements of branching fractions and \CP-violating
asymmetries in $\Bz\to\pip\pim$, $\Kp\pim$, and $\Kp\Km$ decays.
All results are consistent with our previous measurements.
No evidence for \CP violation is observed and
our measurement of ${\cal A}_{K\pi}$
disfavors theoretical models that predict a large 
asymmetry~\cite{PQCD,Charming}.

We are grateful for the 
extraordinary contributions of our \pep2\ colleagues in
achieving the excellent luminosity and machine conditions
that have made this work possible.

\end{document}